\documentclass[3p]{elsarticle}
\usepackage{graphics}
\usepackage{graphicx}
\usepackage{amssymb}

\begin{document}


\begin{frontmatter}

\title{Low temperature resistivity studies of SmB$_6$: 
Observation of two-dimensional variable-range hopping conductivity}

\author[1]{Marianna Batkova\corref{Batkova}}
\cortext[Batkova]{Corresponding author: batkova@saske.sk } 

\author[1]{Ivan Batko}
\author[1]{Slavom\'{i}r Gab\'{a}ni}
\author[1]{Emil Ga\v {z}o}
\author[2]{Elena Konovalova}
\author[2]{Vladimir Filippov}

\address[1]{Institute of Experimental Physics, Slovak Academy of Sciences, Watsonova 47, 040~01~Ko\v {s}ice, Slovakia}
\address[2]{Institute for Problems of Material Science, NASU,  Krzhyzhanovskii str. 3,
    UA-252680~Kiev, Ukraine}

\begin{abstract}

We studied electrical resistance of a single-crystalline SmB$_6$ sample with a focus on the region of the "low-temperature resistivity plateau". Our observations did not show any true saturation of the electrical resistance at temperatures below 3~K down to 70~mK. According to our findings, temperature dependence of the electrical conduction in a certain temperature interval above 70~mK can be decomposed into a temperature-independent term and a temperature-activated term that can be described by variable-range hopping formula for two-dimensional systems, exp[-($T_0/T)^{1/3}$]. Thus, our results indicate importance of  hopping type of electrical transport in the near-surface region of SmB$_6$. 

\end{abstract}


\end{frontmatter}


\biboptions{sort&compress}

\section{Introduction}

	 Heavy fermion semiconductor samarium hexaboride, SmB$_6$,
				attracts attention of researchers for more than half a century. 
	Its fascinating physical properties are very interesting not only because of long-lasting fundamental questions,
				but also from application point of view 	\cite{Wachter93,Allen79,Riseborough00,Yong14,Yong15,Petrushevsky17,Syers15,Wolgast13,Cooley95,Batko93}.   
Electrical resistivity of high-quality SmB$_6$ samples shows a rapid increase with decreasing temperature below 50~K,
				and surprisingly, it shows tendency to saturate at very high residual value $\rho_0$ at the lowest temperatures, below $\approx$ 4~K
				\cite{Batko93,Allen79,Cooley95}, while 											%
the corresponding residual conductivity $\sigma_0$ = 1$/$ $\rho_0$ is less than 
				the minimum metallic conductivity $\sigma_{min}$ defined by Mott \cite{Mott71}.
Thus, in case of metallic conductivity, such unusually low $\sigma_0$ would require a mean free path 
			of itinerant conduction electrons to be less than the interatomic spacing \cite{Allen79,Cooley95}. 
				This unphysical requirement has been a subject of a long-standing controversy whether SmB$_6$ in the ground state is a metal or an insulator, 
				although, in principle, it clearly indicates that electrical transport in SmB$_6$ cannot be assigned to any metallic-type conductivity 
				that would be homogeneous in the whole volume of the material. 

	The strange low-temperature behavior of   SmB$_6$ has been associated with existence of a metallic surface, 
			whereas nontrivial topological surface states \cite{Dzero10,Dzero12,Lu13,Alexandrov13}, 
			as well as ”trivial” polarity-driven ones \cite{Zhu13} have been proposed to exist there.
%
	Valence-fluctuation induced hopping transport \cite{BaBa14} represents another possibility 
			to explain the electrical transport in SmB$_6$ at the lowest temperatures,
			while this scenario moreover predicts enhanced conductivity of the near-surface layer, resembling 
				the typical feature of topological insulators \cite{BaBa14}. 
%
%
In principle, all above mentioned scenarios can coexist,
		but up to now a convincing association of experimentally observed electrical 
		conductivity of SmB$_6$ at lowest temperatures 		with a specific mechanism(s) is still missing.   
A persuasive conclusion about the origin of the resistivity saturation in SmB$_6$ unconditionally requires
		detailed  studies of electrical transport at lowest temperatures allowing to identify corresponding conduction mechanism(s).  
In this work we report electrical resistance studies of single crystal SmB$_6$ focusing on the low-temperature region below 3~K.  



 
\section{Materials and Methods}  

The SmB$_6$ single crystal was grown by the zone-floating method. 
The studied sample (the same one as used  in our previous studies \cite{Batkova06})
 was cut in a bar-shape geometry in the (1~0~0) direction.
Electrical resistivity was measured below 3~K down to 70~mK 
in a home-made dilution $^3$He-$^4$He minirefrigerator.
Four probe ac resistance measurements were done by LR-700
AC Resistance Bridge (Linear Research, USA) in a non-shielded laboratory.
Temperature was determined utilizing a calibrated (Cernox) thermometer (Lake Shore Cryotronics, USA).

\section{Results and discussion}
	
	Previously we studied electrical resistivity of the same SmB$_6$ sample in the temperature interval between 4~K and 100~K \cite{Batkova06}.
As we reported there \cite{Batkova06}, the resistivity below 50~K  shows a
large increase with decreasing temperature
and  has {\em a tendency to saturate} at a high residual value at the lowest
temperatures \cite{Batkova06}. 
Two regions with different  
activation energies were found in the temperature dependence of the resistivity 
  $\rho(T) \propto e^{W/k_{B}T}$ ($k_{B}$ is Boltzmann constant and $W$ is an activation energy), namely   
 $W$  = 5.2 meV for  the interval of 
 6 - 10~K and $W$  = 5.6~meV between 18~K and 30~K \cite{Batkova06}.
Now we have focused on the temperature  interval from  3~K down to 70~mK.
		As can be seen  in Fig.~1, the obtained data show no saturation 
in the investigated temperature range (despite to indication based on the previous measurements above 4~K \cite{Batkova06}).
Instead of this, two inflex points in the $\rho(T)$ curve have been found in this temperature interval, 
which can be considered as a sign for different dominating transport regimes below and above 1~K.
(Such resistivity behaviour qualitatively resembles one of another  single crystalline sample of SmB$_6$ 
prepared by the same technology process \cite{Batko93}.)

			%
		\begin{figure}[!t]
			\center{
				\resizebox{1.00\columnwidth}{!}{%
  				\includegraphics{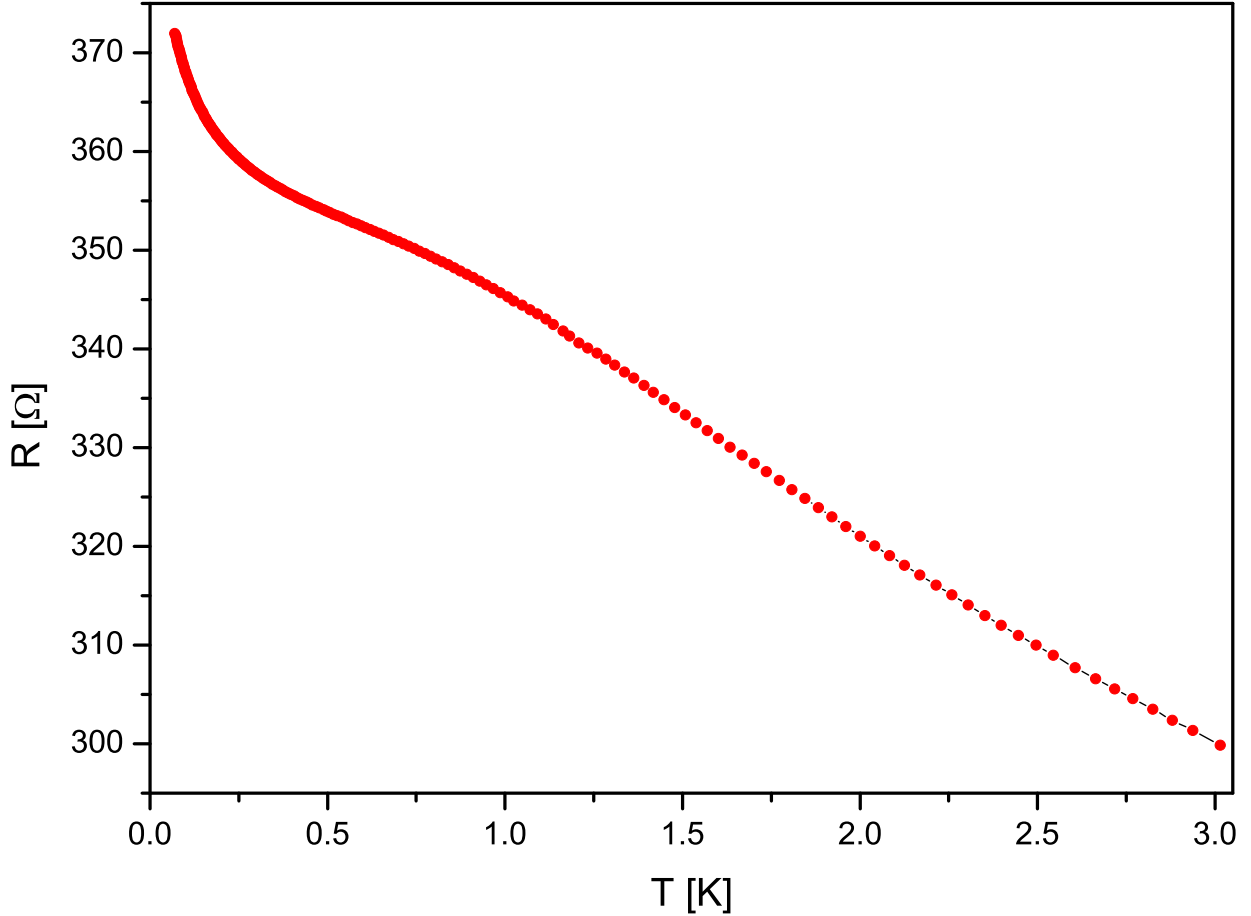}
        }
    	}
	 			\caption{Temperature dependence of the resistance for the single-crystalline sample of SmB$_6$ between 70~mK and 3~K. The resistivity calculated as for a "homogeneously" conductive sample is approximately 20~$\Omega$cm at 3~K. }
			\label{Fig1}
		\end{figure}
			%

As reported by many authors before, the low-temperature dependence of the electrical resistance, $R(T)$,  can be 
in case of SmB$_6$ described by the two-channel model \cite{Petrushevsky17,Syers15,Wolgast13,Cooley95,Batko93}, e.g. in the form  
\begin{equation}
	R(T)^{-1} = R_{n}^{-1} + R_{a}^{-1} \times \exp(-W_{a}/kT),
\label{2ch}	
\end{equation} 
 %
%
%
%
where $W_a$ is activation energy associated with temperature activated transport of the bulk, 
and parameters $R_{n}^{-1}$ and $R_{a}^{-1}$ are constants characterizing temperature non-activated and temperature activated
contribution to the conduction, 
and are usually associated with metallic nature of the surface and insulating bulk, respectively. 
Our experimental data, depicted in Fig.~\ref{Fig1}, also indicate presence of temperature activated transport at lowest temperatures, below 0.5~K.
Therefore we have made attempts to fit the  data below 0.5~K using Eq.~\ref{2ch}. 
However, we have found out that this type of regression formula is not suitable, and another model 
of electrical transport needs to be considered to describe the data.

Another mechanism of temperature activated electrical conductivity, typical for heavily doped and amorphous
semiconductors at low temperatures, is variable-range hopping (VRH) between localized states in
the vicinity of the Fermi level \cite{Mott71, Shklovskij84}. 
In the most general form the resistivity in such case is given by
\begin{equation}
	\rho(T) \propto \exp{[(T_0/T)^p]}, 
\label{vrh}	
\end{equation} 
where $p$ depends on the dimensionality, $d$, 
of the system and the energy distribution of the density of localized states  
in the vicinity of the Fermi level. 
If the density of localized states near the Fermi level is
constant or depends only slowly on energy, then  $p=(d+1)^{-1}$ \cite{Shklovskij84},
thus for $d=3$  the
temperature dependence of the resistivity follows the
Mott's law, i.e. $\rho(T) \propto \exp{[(T_0/T)^{1/4}]}$, 
and in two dimensional systems 
$\rho(T) \propto \exp{[(T_0/T)^{1/3}]}$.
Supposing that the electrical conduction of the studied SmB$_6$ sample can be  described by the two-channel model, 
where one channel represents temperature non-activated contribution, 
and the second one is of a VRH type 
then it should be true that
\begin{equation}
	R(T)^{-1} = R_{n}^{-1} + R_{3D}^{-1} \times \exp{[-(T_{3D}/T)^{1/4}]},
\label{vrh3d}	
\end{equation}
where  parameters $R_{3D}^{-1}$ and $T_{3D}$ are constants characterizing
VRH contribution to the bulk conduction, 
and $R_{n}^{-1}$ has similar meaning as in Eq.~\ref{2ch} (and in Eq.~\ref{vrh2d} below).
It has been noticed that  (significantly) increased concentration of lattice imperfections,   
which  may play role of hopping centers, is expected in the near-surface region of SmB$_6$,
and therefore a strong enhancement of the conductivity in this region is expected \cite{BaBa14}. 
Supposing that this near-surface region can be treated as a 2D layer, the resulting conduction
can be written in the form
\begin{equation}
	R(T)^{-1} = R_{n}^{-1} + R_{2D}^{-1} \times \exp{[-(T_{2D}/T)^{1/3}]},
\label{vrh2d}	
\end{equation} 
where   $R_{2D}^{-1}$ and $T_{2D}$ are constants characterizing 
 VRH transport in this near-surface region.

\begin{figure}[!t]
			\center{
				\resizebox{1.00\columnwidth}{!}{%
  				\includegraphics{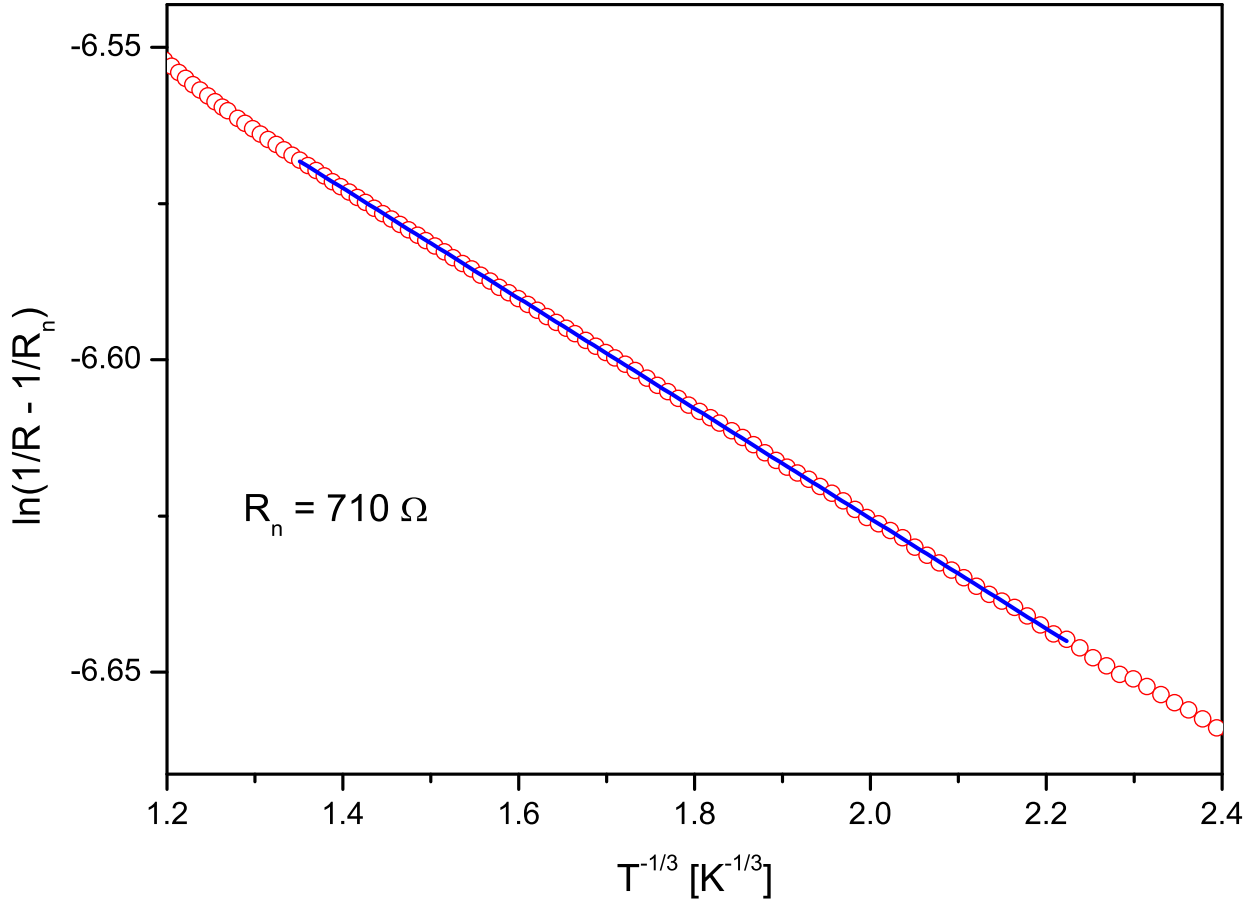}
        }
    	}
	 			\caption{Plot of the resistance of the SmB$_6$ single crystal
			in coordinates $ln(1/R - 1/R_n)$ {\em versus} $T^{-1/3}$ 
			($R_n = 710~\Omega$) demonstrating that temperature activated transport below 0.45~K ($T^{-1/3} > 1.3$~K$^{-1/3}$) is governed by two-dimensional VRH.
			The blue solid line represents the fit of the experimental  data (red circles) by means of Eq. \ref{vrh2d} in the temperature range of 0.09 - 0.4~K  (see text).}
			\label{Fig2}
		\end{figure}

Attempts to fit the experimental data at lowest temperatures 
using the regression formula expressed by Eq.~\ref{vrh3d} did not provided satisfactory result,
thus indicating absence of VRH type of transport in the bulk of the sample at lowest temperatures.
On the other hand, performed numerical analysis revealed that regression formula in the form of Eq.~\ref{vrh2d} is
suitable for the fitting. 
%
%
As can be seen in Fig.~\ref{Fig2}, where  the resistance data are plotted 
			in coordinates $ln(1/R - 1/R_n)$ {\em versus} $T^{-1/3}$, 
			Eq.~\ref{vrh2d} provides excellent description of the data at temperatures below 0.45~K. 
The resulting fitting parameters $R_n$ = 710~$\Omega$, $R_{2D}$ = 632~$\Omega$  and $T_{2D} = 6.85 \times 10^{-4}$~K were obtained based on the fit of the data from the temperature interval of 0.09 - 0.4~K.
In accordance with the discussion above we take this as evidence that  VRH transport
in the near-surface region, being a consequence of significantly higher concentration of lattice imperfections in this region, is dominating temperature activated conduction mechanism in SmB$_6$ below $\approx$0.45~K.

%
		%

Now, let us focus  on the data above 1~K.
Several authors modeled electrical resistivity data of SmB$_6$ at temperatures above 2~K by a formula quivalent to Eq.~\ref{2ch} \cite{Petrushevsky17,Syers15,Wolgast13,Cooley95,Batko93}.   
However, performing numerical analysis of our data we have found out that although this formula is appropriate for a rough description
 of the experimental data at temperatures approching 3~K, there is a systematic deviation of the fit by this type of formula 
from the experimental data. 
This can indicate that the temperature dependent 
two-dimensional VRH, which dominates at the lowest temperatures, is not negligible 
neither at temperatures approaching 3~K, so 
 consideration of two-dimensional VRH is still needed to describe the electrical transport in SmB$_6$  above 1~K.
Therefore we analyzed the conduction data above 1~K using a combination of Eq.~\ref{2ch} and Eq.~\ref{vrh2d} 
in the form 
\begin{equation}
R(T)^{-1} = R_{n}^{-1} + R_{2D}^{-1} \times \exp{[-(T_{2D}/T)^{1/3}]} + R_{a}^{-1} \times \exp(-W_{a}/kT),
\label{comb}	
\end{equation}

\begin{figure}[!t]
			\center{
				\resizebox{1.00\columnwidth}{!}{%
  				\includegraphics{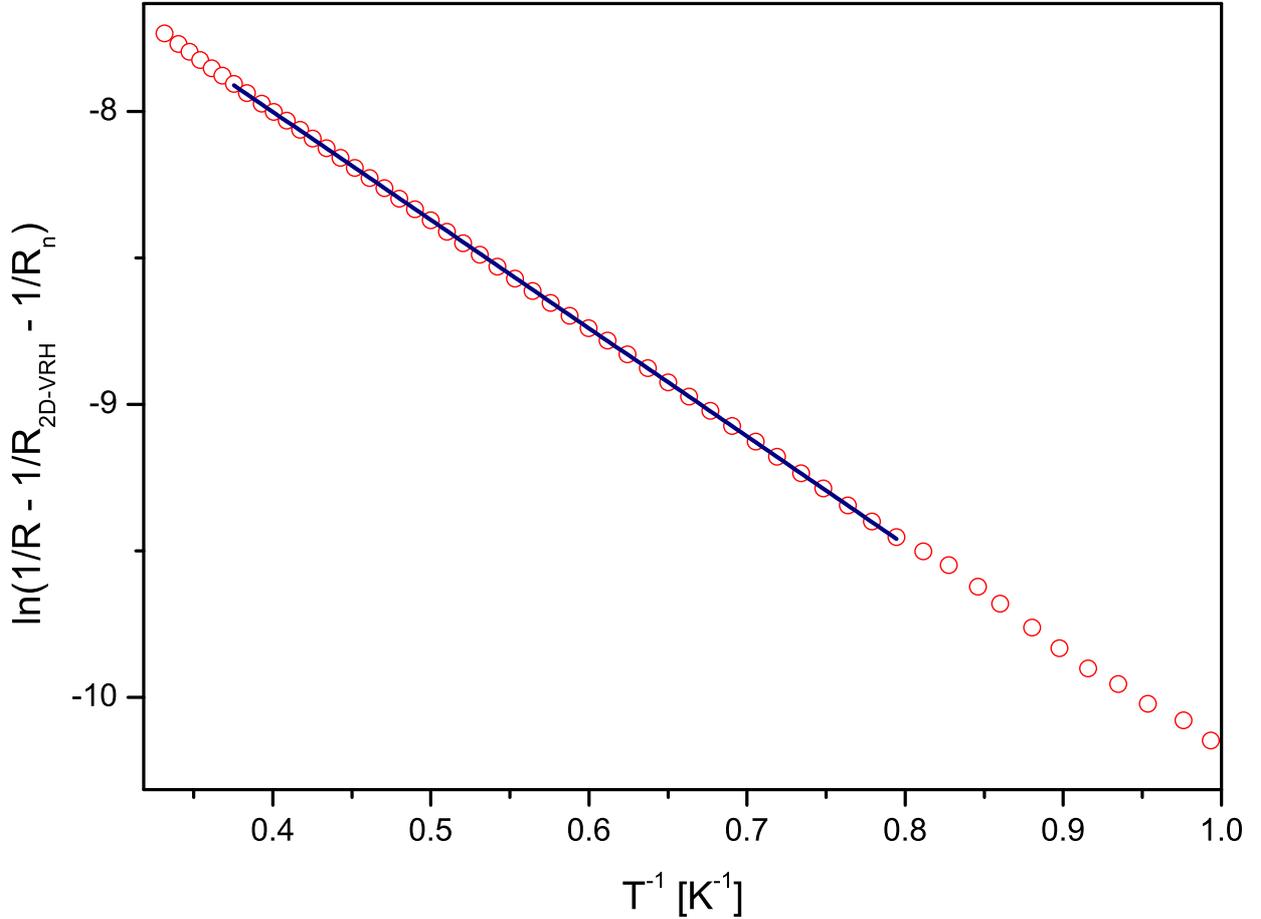}
        }
    	}
	 			\caption{Plot of logarithm of the conduction data 
after subtraction of two-dimensional VRH contribution, 
$R^{-1}_{2D-VRH} = R_{2D}^{-1} \times \exp{[-(T_{2D}/T)^{1/3}]}$, 
and temperature non-activated conduction,  $R_{n}^{-1}$, 
as determined from the fit between 0.09~K and 0.4~K (see text). 
Activation energy $W_a = 0.32$~meV was determined from the slope of the linear fit (blue solid line)
of the data in the range between 1.25~K and 2.5~K. 
}
			\label{Fig3}
		\end{figure}

Indeed,  taking into account not only  the contribution due to the temperature activated conduction in the bulk (and the temperature non-activated term),  but also the term due to the two-dimensional VRH, has lead to finding a proper fit 
of the data in the temperature range of 1.25 -2.5~K.
It is demonstrated in Fig.~\ref{Fig3}, where logarithm of the conduction data 
after subtraction of  two-dimensional VRH contribution, $R_{2D-VRH}^{-1} = R_{2D}^{-1} \times \exp{[-(T_{2D}/T)^{1/3}]}$, 
and temperature non-activated conduction,  $R_{n}^{-1}$, 
as determined from the data between 0.09~K and 0.4~K 
is plotted versus $T^{-1}$.
The  activation energy determined  as the slope of the linear fit 
of the data from Fig.~\ref{Fig3} in the range between 1.25~K and 2.5~K is $W_a = 0.32$~meV.
We associate this energy with the impurity-to-band activation energy of impurities (and lattice imperfections) forming the impurity band in the forbidden gap of SmB$_6$.
(Note that $W_a$  is nearly one order of magnitude less than the energy of activation process governed by the forbidden gap, $W$,
 determined for this sample previously \cite{Batkova06}.) 
 
%

			%

			Analysis performed above indicates that electrical transport in  SmB$_6$ above 1~K can be adequately explained considering
two temperature activated terms, which we predominantly associate with the conduction of the bulk,  
$R_{a}^{-1} \times \exp(-W_{a}/kT)$, and the two-dimensional VRH conduction in the near-surface region, $R_{2D}^{-1} \times \exp{[-(T_{2D}/T)^{1/3}]}$. 
The additional temperature non-activated term ($R_{n}^{-1}$) can be in accordance
with existing models ascribed to metallic surface states, which can be either trivial  \cite{Zhu13} or topologically protected \cite{Dzero10,Dzero12,Lu13,Alexandrov13}, or alternatively it can be associated with valence-fluctuation induced hopping transport \cite{BaBa14} .

\section{Conclusions}
Performed electrical resistance studies  of the single-crystalline SmB$_6$ sample reveal that 
there is no true saturation of the electrical resistance down to temperature as low as 70~mK.  
The electrical conduction in a certain temperature interval above 70~mK can be described by the two-channel model 
with a temperature independent (metallic-like) term and temperature activated term corresponding to  variable-range hopping conduction 
in two-dimensional systems, which can be associated with
an enhanced conduction in the near-surface region. 
The three-channel model, considering moreover temperature activated conduction from the bulk,
 is needed to describe the experimental data above 1~K.  
Our results moreover indicate that precision resistance measurements down to millikelvin temperatures are needed   
for reliable conclusions about the ground state of SmB$_6$.

\section*{Acknowledgement}
This work was supported by the Slovak Scientific Agency VEGA (grant No.~2-0015-17) and by the Slovak Research and Development Agency (contract No. APVV-0605-14).

%
%

\section*{References}


\end{document}